%% file: interact2023-digital-modeling-for-everyone.tex
\documentclass[runningheads]{llncs}
\usepackage[crop,final,llncs]{espositollncs}
\conference{19th IFIP TC 13 Conference on Human-Computer Interaction (INTERACT 2023)}
\llncsdoi{10.1007/978-3-031-42293-5_11}
\llncs{J. Abdelnour Nocera et al. (Eds.): INTERACT 2023, LNCS 14145, 2023}{133}

\usepackage{graphicx}

\input{meta/packages.tex}
\usepackage[capitalise]{cleveref}

\usepackage[normalem]{ulem}

\input{meta/acronyms.tex}

\begin{document}

\title{Digital Modeling for Everyone: Exploring How Novices Approach Voice-Based 3D Modeling}
\titlerunning{Digital Modeling for Everyone}
\input{meta/authors}

\maketitle

\begin{abstract}
Manufacturing tools like 3D printers have become accessible to the wider society, making the promise of digital fabrication for everyone seemingly reachable. While the actual manufacturing process is largely automated today, users still require knowledge of complex design applications to produce ready-designed objects and adapt them to their needs or design new objects from scratch. To lower the barrier to the design and customization of personalized 3D models, we explored novice mental models in voice-based 3D modeling by conducting a high-fidelity Wizard of Oz study with 22 participants. We performed a thematic analysis of the collected data to understand how the mental model of novices translates into voice-based 3D modeling. We conclude with design implications for voice assistants. For example, they have to: deal with vague, incomplete and wrong commands; provide a set of straightforward commands to shape simple and composite objects; and offer different strategies to select 3D objects.

\end{abstract}

\input{content/introduction}
\input{content/relatedwork}
\input{content/methodology}
\input{content/results}
\input{content/discussion}
\input{content/limitations}
\input{content/conclusion}

\subsubsection{Acknowledgements}
\input{content/acknowledgments}

\bibliographystyle{splncs04}
\bibliography{bib}
\end{document}

%% file: meta/packages.tex
\usepackage{siunitx}
\usepackage{calculator}
\usepackage{booktabs}
\usepackage{fmtcount}
\usepackage{float}

\usepackage[figuresright]{rotating}
\usepackage{subcaption}

\usepackage{tikz}
\usetikzlibrary{chains,arrows}
\usepackage{fontawesome}

\usepackage{enumerate} 
\usepackage[inline]{enumitem} 
\usepackage{tabularx} 

\usepackage{hyperref}

%% file: meta/acronyms.tex
\usepackage[nolist,nohyperlinks]{acronym}

\begin{acronym}[UMLX]
    \acro{GDPR}{General Data Protection Regulation}
	\acro{XAI}{eXplainable Artificial Intelligence}
	\acro{HMD}{head-mounted display}
	\acro{GUI}{graphical user interface}
	\acro{HUD}{head-up display}
	\acro{TCT}{task-completion time}
	\acro{SVM}{support vector machine}
	\acro{EMM}{estimated marginal mean}
	\acro{AR}{Augmented Reality}
	\acro{VR}{Virtual Reality}
	\acro{RTLX}{Raw Nasa-TLX}
	\acro{WUI}{Walking User Interface}
	\acro{CSCW}{Computer-Supported Cooperative Work}
	\acro{HCI}{Human-Computer Interaction}
	\acro{ABI}{Around-Body Interaction}
	\acro{SLAM}{Simultaneous Location and Mapping}
	\acro{ROM}{range of motion}
	\acro{EMG}{electromyography}
	\acro{STS}{System Trust Scale}
	\acro{IPQ}{System Trust Scale}
	\acro{TLX}{NASA Task Load Index}
	\acro{ART}{Aligned Rank Transform}
	\acro{CNC}{Computerized Numerical Control}
	
    \acro{ML}{Machine Learning}
    \acro{AI}{Artificial Intelligence}
    \acro{NLP}{Natural Language Processing}
    \acro{NLU}{Natural Language Understanding}
    \acro{SVM}{Support-Vector Machine}
    \acro{VA}{Voice Assistant}
    \acro{CA}{Conversational Agent}
    \acro{HCI}{Human-Computer Interaction}
    \acro{HCAI}{Human-Centred Artificial Intelligence}
    \acro{WoZ}{Wizard of Oz}
    \acro{HCI}{Human-Computer Interaction}
    \acro{IT}{Information Technology}
    \acro{BERT}{Bidirectional Encoder Representations from Transformers}
    \acro{T5}{Text-To-Text Transfer Transformer}
    \acro{C4}{Colossal Clean Crawled Corpus}
    \acro{STM}{Speech Transformer Model}
    \acro{ASR}{Automatic Speech Recognition}
    \acro{TTS}{Text to Speech}
    \acro{LAS}{Listen, Attend and Spell}
    \acro{SEQ2SEQ}{Sequence to Sequence}
    \acro{AR}{Augmented Reality}
    \acro{VR}{Virtual Reality}
    \acro{EUD}{End-User Development}
    \acro{API}{Application Programming Interface}
    \acro{REST}{Representational State Transfer}
    \acro{RNN}{Recurrent Neural Network}
    \acro{LSTM}{Long Short-Term Memory}
    \acro{CAD}{Computer-Aided Design}
    \acro{GUI}{Graphical User Interface}
    \acro{UI}{User Interface}
    \acro{WIMP}{Window, Icon, Menu, Pointer}
    \acro{CNC}{Computer Numerical Control}
    \acro{SRQ}{sub-research question}
\end{acronym}

%% file: meta/authors.tex
\author{Giuseppe Desolda\inst{1}\orcidID{0000-0001-9894-2116} \and
Andrea Esposito\inst{1}\orcidID{0000-0002-9536-3087} \and
Florian M\"uller\inst{2}\orcidID{0000-0002-9621-6214}
\and
Sebastian Feger\inst{2}\orcidID{0000-0002-0287-0945}%
}
\authorrunning{G. Desolda et al.}
%
\institute{
	Department of Computer Science, University of Bari Aldo Moro, Bari, Italy\\
\email{\{giuseppe.desolda, andrea.esposito\}@uniba.it}
\and
    LMU Munich, Munich, Germany \\
\email{\{florian.mueller, sebastian.feger\}@um.ifi.lmu.de}
}

%% file: content/introduction.tex
\section{Introduction}
\label{sec:introduction}

The digital fabrication revolution aims to democratize the way people create tangible objects \cite{gershenfeld2012make}. With the widespread availability of 3D printing together with many other digital fabrication technologies such as laser cutters or \ac{CNC} routers, end users are moving from passive consumers to active producers. While the actual manufacturing process is largely automated today, users are still required to have a profound knowledge of complex 3D modeling applications, when they adapt models to their needs or even design new objects from scratch \cite{xue2010command}. Thus, even if the introduction of technologies such as 3D printers has revolutionized the hobbyist community, lowering the barrier of entry to manufacturing even for novices (who can now put their hands in the process of creating artifacts without relying on third parties), we argue that the design of the 3D objects to be manufactured still requires a high level of knowledge and expertise. 

These limitations have pushed researchers to investigate natural interaction techniques to simplify 3D modeling tools \cite{niu2022multimodal}. For example, research explored gestures \cite{Thakur2015,VULETIC2021102609}, virtual/augmented reality \cite{STARK2010179,Feeman2018}, eye tracking \cite{JOWERS2013923,Yoon2008}, brain-computer interface \cite{Huang2017,SREESHANKAR201451} and their combination \cite{Nanjundaswamy2013,friedrich2020combining,khan20193d,KHAN2019102847} as a multimodal approach. However, their adoption is reserved for technical users and it is strongly limited by hardware costs and excessive size/weight that can make the users easily fatigued \cite{niu2022multimodal}. As another possible solution, voice-based interaction has been explored, to both integrate the traditional \ac{GUI} interface (e.g., to enable shortcuts via voice commands) \cite{Voice2CAD,xue2010command}) or as the primary interaction paradigm (e.g., see \cite{xue2009natural,KOU2010545,plumed2021voice}). Although voice-based interaction requires only a microphone, it does not yet provide adequate digital modeling support for everyone: existing solutions either do not consider final users at all \cite{xue2010command,xue2009natural}, or only target 3D experts \cite{KOU2010545,Making_A_Scene,KHAN2019102847,plumed2021voice}, and novices are not considered potential target beneficiaries of the proposed innovations.

To lower the barrier to the design and customization of personalized 3D models by exploiting the potential of voice-based interaction, this study aims to understand how the mental model of novices translates into voice-based 3D modeling. We conducted a high-fidelity \ac{WoZ} study to elicit novices' mental model, for example, their expectation, beliefs, needs, and abilities. We recruited a total of 22 participants without skills in 3D modeling, who performed 14 tasks revolving around some basic concepts of 3D modeling like the creation of objects, the manipulation of objects (e.g., scaling, rotating, and/or moving objects), and the creation of composite objects. All the \ac{WoZ} sessions' recordings were analyzed through thematic analysis. The findings of the study have been distilled in the form of lessons learned. For example, we found that: voice assistants must manage the corrections the novices do during and after the commands; deal with vague and incomplete commands; consider the prior novices' knowledge; provide only a simplified set of operations for creating simple and composite 3D objects; design a workflow similar to what novices would do if they were building real objects; understand chained commands; understand commands that are relative to the users’ point of view. 

The contribution of this paper is two-fold. First, we report the results of our \ac{WoZ} study presenting the themes that emerged from the thematic analysis. Second, based on these results, we provide a set of design implications for the future design of voice-based interaction paradigms for 3D modeling for novices.

%% file: content/relatedwork.tex
\section{Background and Related Work}
\label{sec:relatedwork}

This study revolves around the concept of voice-based 3D modeling as a key factor for enabling the democratization of digital fabrication. This section starts by illustrating some of the existing solutions based on natural interaction that try to address the complexity of 3D modeling (\cref{subsec:address-complexity-of-cad}). Next, we provide an overview of the requirements for interacting with voice assistants (\cref{subsec:interacting-with-voice-assistants}). Finally, we provide a brief summary of the motivation of this study and introduce the research question that guided our work (\cref{subsec:summary_research_question}).

\subsection{Addressing the Complexity of 3D modeling}\label{subsec:address-complexity-of-cad}

To mitigate the issues of traditional \ac{GUI}-based \ac{CAD} systems, researchers explored natural interaction paradigms like eye tracking \cite{JOWERS2013923,Yoon2008}, brain-computer interface \cite{Huang2017,SREESHANKAR201451}, gestures \cite{Thakur2015,VULETIC2021102609}, virtual/augmented reality \cite{STARK2010179,Feeman2018} and their combination \cite{friedrich2020combining,khan20193d,KHAN2019102847} as a multimodal approach for 3D modeling. The goal of natural interactions with \ac{CAD} systems is to increase their \emph{usability} for both expert users and, especially, novice users. Specifically, they aim to: \begin{enumerate*}[label=\roman*)]
    \item reduce the learning curve of the system;
    \item allow a more intuitive interaction process;
    \item enhance the design abilities of the designers
\end{enumerate*}
\cite{niu2022multimodal}.

An example of a multimodal system is ``3D Palette'' by Billinghurst et al.: a mix of tablet and pen inputs, electromagnetic sensors and voice commands are used to support the digital design process \cite{Billinghurst1997}. Similarly, Nanjundaswamy et al\@. explored a mix of gesture-based interaction, speech recognition, and brain-computer interfaces to reduce the initial learning curve of the design system \cite{Nanjundaswamy2013}. A complete overview of the multimodal solutions for \ac{CAD} is reported by Niu et al.~\cite{niu2022multimodal}. Despite these potential benefits, such multimodal techniques require the adoption of specialized hardware (e.g., depth-sensing cameras for gesture recognition, headsets to recognize brain signals), which use can be limited by their prices, sizes, weight, and complexity of use \cite{Nanjundaswamy2013}. Thus, it is still hard for novice users to really adopt them in real and daily contexts \cite{niu2022multimodal}. 

To overcome these limitations, researchers also investigated voice-based interaction because of its intuitive nature and the simplicity of the required hardware, i.e., a microphone, which nowadays is embedded in any laptop, tablet, or webcam \cite{seaborn2021voice}. Furthermore, considering the ubiquity of smartphones and the rise of \acs{AR} and \acs{VR} glasses, voice-based interaction can be generalized to technologies where other interaction modalities are not available options. Attempts of integrating voice-based interaction to \ac{CAD} systems date as back as 1985 \cite{Samad1985}. A more recent work suggests the use of voice commands to allow users to either quickly search commands by simply stating their intention \cite{xue2010command,Voice2CAD}, or to annotate 3D models \cite{plumed2021voice}. Systems, where the entire modeling process is carried out by voice commands, have also been explored. An example is the solution presented by Kou and Tan, where voice commands related to a \ac{CAD}-specific lexicon and grammar are understood by a context-aware algorithm \cite{Kou2008}. A similar example was proposed by Xue et al., which improves the previous solution by allowing free-form sentences in \cite{xue2009natural}. Another example of a fully-working system is the one presented by Grigor et al.: it follows the same ideas as the previous ones but uses \ac{AI} to understand the users' inputs, thus allowing for more freedom in the commands, \cite{grigorvoice}. Similarly, Kou et al\@. proposed a flexible voice-enabled \ac{CAD} system, where users are no longer constrained by predefined commands by exploiting a knowledge-guided approach to infer the semantics of voice input \cite{KOU2010545}.

Among all the previous examples, it must be highlighted that the design of their paradigm was made without any kind of involvement of the final users \cite{Samad1985,xue2010command,Voice2CAD,Kou2008} or by solely involving experts in the final testing phase \cite{grigorvoice}. For example, the study by Nanjundaswamy et al\@. evaluates a multimodal system using gestures, speech and a brain-computer interface by involving a group of five skilled people \cite{Nanjundaswamy2013}. Similarly, Khan et al\@. involve a total of 41 skilled users from an architecture or engineering background to elicit the requirements of a \ac{CAD} system based on gestures and speech commands \cite{KHAN2019102847}. As another example, Vyas et al\@. test the usability of a speech-based \ac{CAD} system involving 6 students with backgrounds in engineering, architecture and visualization \cite{Making_A_Scene}.

The work proposed by Cuadra et al\@. investigated how novices use voice assistants to design 3D objects \cite{Cuadra2021}. They performed a \ac{WoZ} study to compare voice assistants with and without the use of a video channel showing the design in progress, investigating how the two approaches impact users' accuracy and satisfaction. Cuadra et al\@. validate the idea of using voice assistants, as participants are more satisfied with their objects and suffer less from cognitive overload when the design process is supported by video, but it does not provide any insight on the mental model of novices approaching the digital modeling task \cite{Cuadra2021}.

\subsection{Interacting with Voice Assistants}
\label{subsec:interacting-with-voice-assistants}

The first solution of voice interaction implementing speech recognition dates as back as 1952, when Davis et al\@. proposed a prototype able to recognize digits \cite{davis1952automatic}. In recent years, the evolution of machine learning and \ac{AI} fostered the spreading of powerful commercial voice assistants, often based on deep neural networks trained on a plethora of data. 
However, such powerful speech recognition models alone are not sufficient to build an effective voice assistant, since the interaction with such systems must be considered in the design of the whole system \cite{Moore2019}. This need, together with the growing availability of commercial voice assistants, has fostered a sharp uptick of studies on user interaction with voice assistants \cite{seaborn2021voice}. Aspects like the cues that drive the conversation \cite{Vtyurina2018}, the properties that a voice assistant should have \cite{Volkel2021}, the user's mental model \cite{GRIMES2021113515}, emotions felt during the conversation \cite{James2018}, conversational design patterns \cite{Moore2019} have been investigated. In addition, solutions to design and evaluate interaction with voice assistants are beginning to be proposed (see, for example, \cite{Moore2019,Langevin2021,Murad2019,nowacki2020improving,murad2021finding,jain2018evaluating,Volkel2021}). Careful consideration of these design aspects gains importance when voice assistants aim to simplify challenging or technical operations (e.g., see \cite{Braun2019}). Since 3D modeling represents such a demanding task for novices, the elicitation of the novices' mental model is crucial to lower the barrier for 3D modeling.

\subsection{Summary and Research Question} 
\label{subsec:summary_research_question}

The analysis of the literature highlights that to simplify the 3D modeling, often the existing solutions are based on multimodal techniques such as gestures, eye tracking, or brain-computer interfaces; however, their adoption in real contexts is strongly limited by the adoption of specialized hardware and, overall, they target technical users. 

Voice interaction seems a promising paradigm that can overcome the limitations of multimodal solutions, but the existing voice-based solutions are still lacking for three important reasons:
\begin{enumerate*}[label=\roman*)]
    \item users are often not considered throughout the design phase, or they are only involved too late in testing phases;
    \item to the best of our knowledge, novices are never considered as target users;
    \item the voice-based interaction is built on top of the existing \ac{CAD} systems (and their complexity), instead of designing from scratch the voice paradigm and the whole system.
\end{enumerate*}

Considering these limitations, to really democratize digital fabrication considering novices, users should be able to access 3D modeling tools even without special skills. All these motivations pushed us to explore novices' mental model in voice-based 3D modeling, in order to reduce the cost of their entry in the digital fabrication era. This is an aspect that has never been explored before and that deserves attention to really democratize digital fabrication. Therefore, our work addresses the following research question: \textbf{How does the mental model of novices translate into voice-based 3D modeling?}

%% file: content/methodology.tex
\section{Method}
\label{sec:methodology}

To answer our research question, we performed a high-fidelity \acf{WoZ} study \cite{Sharp2019} because it has been proven successful in eliciting the user's mental model for voice-based interaction (e.g., see \cite{Fialho2015,Vtyurina2018,medhi2017you,Cuadra2021}). Then, we carried out an inductive thematic analysis \cite{Braun2006} on the qualitative data, i.e., the transcriptions of the \ac{WoZ} sessions and the answers of the participants to the open questions. 

\subsection{Participants}\label{subsec:woo-participants}
A total of 22 participants (F=15, M=7) have been recruited through convenience sampling \cite{Etikan2016} on the social circles of the authors of this article. This number of participants is in line with other similar studies (e.g., see \cite{Vtyurina2018,Lee2008}). Half of the participants were Italians while the other half were Germans. Their mean age was 24.1 years ($\sigma$ = 3.7, min = 21, max = 34). The entire study was performed in English so as not to have results related to specific languages, which is out of the scope of this study. To ensure that the collected data is not biased toward knowledgeable users, we only recruited participants without any kind of experience with 3D modeling. Regarding the participants' level of education, around 45.45\% already have a High School Diploma or a German A-level, 36.36\% have a Bachelor's Degree, 13.64\% have a Master's Degree, and only one participant (representing the remaining 4.55\%) has not provided any information. Most participants (15 out of 22) do not have a STEM education, while 6 of the remaining 7 do not have any computational thinking skills, as they studied or worked in non-IT scientific fields (e.g., pharmaceutical and nutrition sciences). Regarding the participants' skills, they had an average level of IT knowledge ($\bar{x}$ = 6.5/10; $\sigma$ = 2.1), a medium-low level of knowledge of voice assistants ($\bar{x}$ = 3.1/10; $\sigma$ = 2.0) and very low knowledge of 3D modeling ($\bar{x}$ = 1.6/10; $\sigma$ = 1.1).

\subsection{Tasks}\label{subsec:woo-tasks}
\begin{figure}[t!]
	\centering
	\begin{subfigure}[t]{0.32\linewidth}
		\centering
		\includegraphics[width=\linewidth]{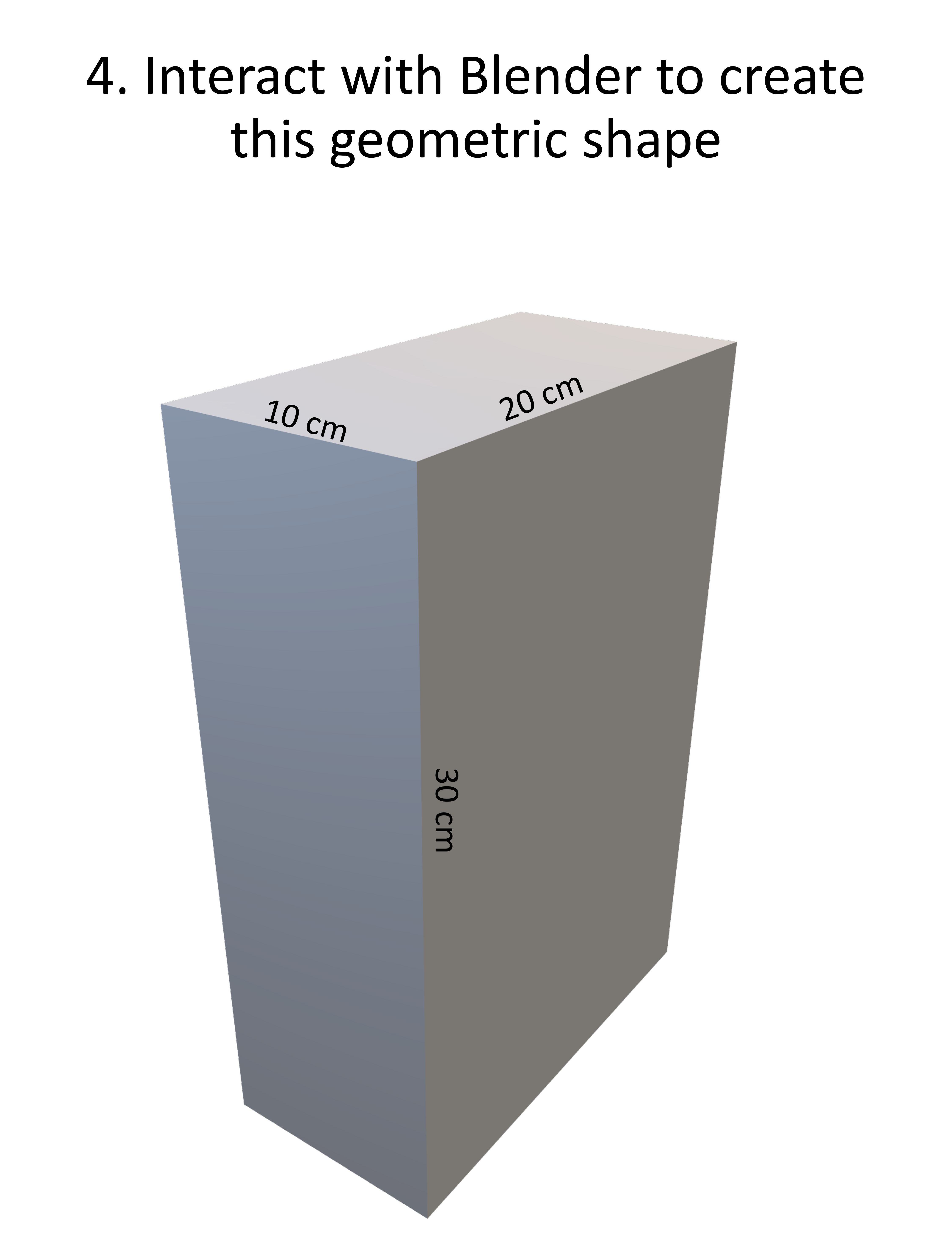}
		\caption{Creation of a simple object}
		\label{fig:create-task}
	\end{subfigure}
	\begin{subfigure}[t]{0.32\linewidth}
		\centering
		\includegraphics[width=\linewidth]{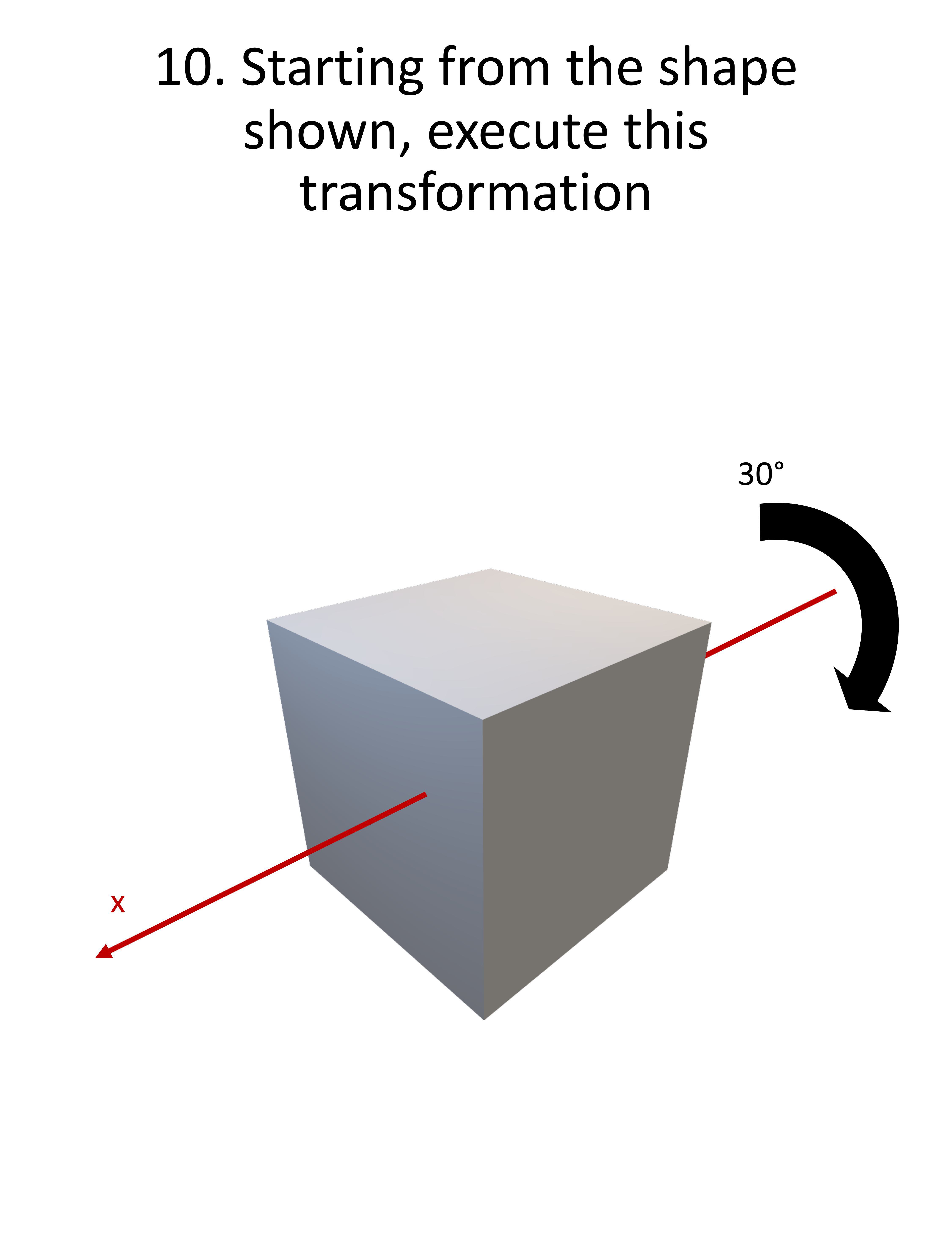}
		\caption{Transformation of an object (rotation)}
		\label{fig:rotate-task}
	\end{subfigure}
	\begin{subfigure}[t]{0.32\linewidth}
		\centering
		\includegraphics[width=\linewidth]{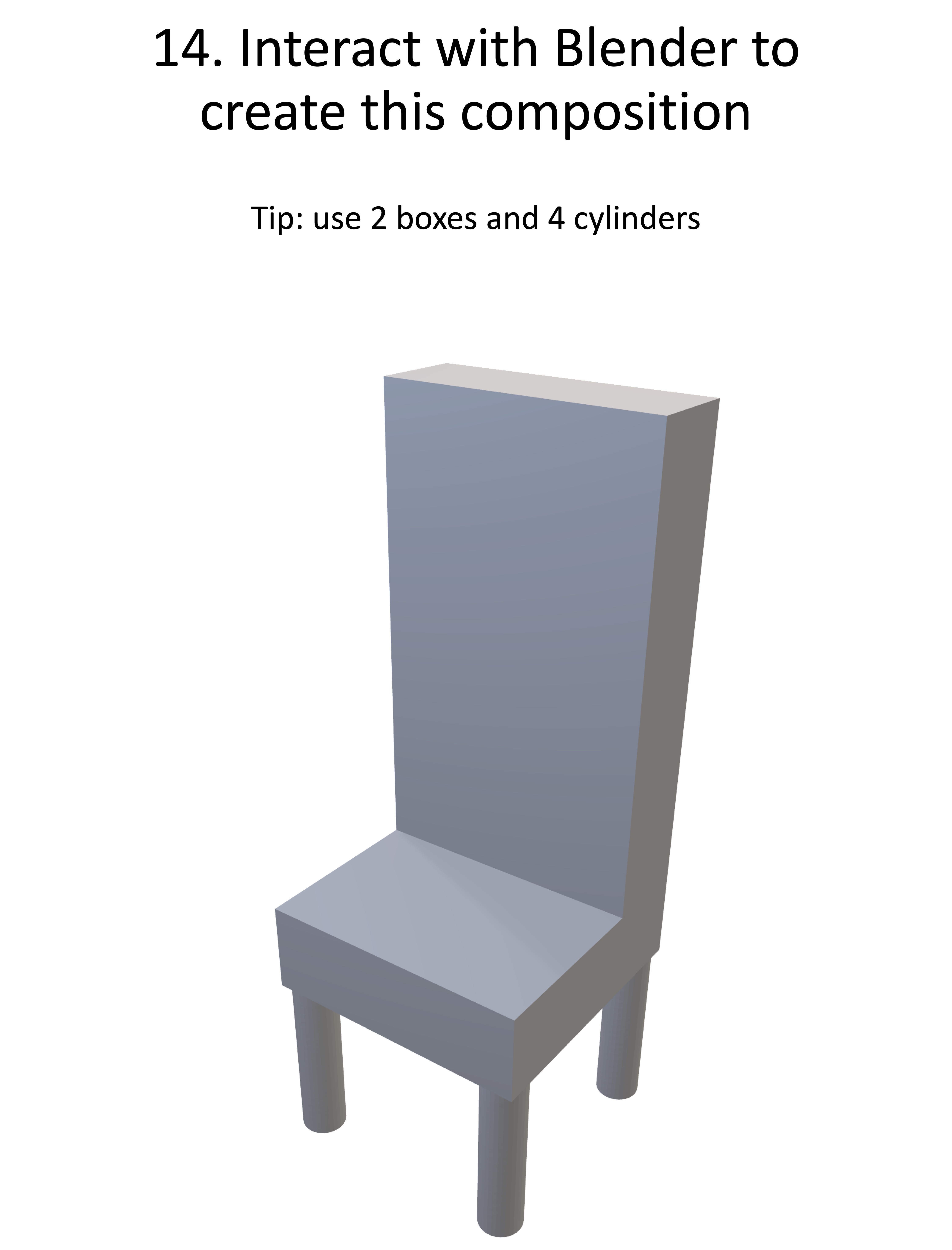}
		\caption{Creation of a composite object}
		\label{fig:chair-task}
	\end{subfigure}
	\caption{Examples of graphical tasks: a brief prompt is reported on top of each task and below a diagram shows the participants the 3D object to create (\protect\subref{fig:create-task}, \protect\subref{fig:chair-task}) or the transformation to be performed (\protect\subref{fig:rotate-task}).}
	\label{fig:example-tasks}
\end{figure}

A total of 14 tasks have been designed by two authors of this paper, both experts in 3D modeling, taking into account the most common and useful activities that are required to create simple and composite 3D objects. The resulting tasks revolve around basic concepts of 3D modeling, like the creation of simple objects, the manipulation of objects (e.g., scaling, rotating, and/or moving objects), and the creation of composite geometries. The details of the tasks are reported in 
the task table in the attached appendix
(the list of all the graphical tasks is available in the attached appendix, sub-folder \textit{tasks}). To reduce the impact of the primer effect \cite{Etikan2016} that providing a textual description of a task would have on the participants, we chose to provide the participants with graphical tasks: each task is composed of a brief prompt and a diagram showing the participants a 3D object or a 3D transformation that should be recreated (an example of graphical tasks is provided in \cref{fig:example-tasks}). The representations chosen for each task were validated during a pilot study with 3 novices that were not considered in the final \ac{WoZ} study.

\subsection{Apparatus}\label{subsec:apparatus}

We carried out the \ac{WoZ} study remotely by using Zoom\footnote{\url{https://zoom.us}}. Four researchers have been involved: two Italians acted respectively as conductors and wizards for the Italian participants, while two German researchers acted as conductors and wizards for the German participants. In both groups, researchers switched roles to minimize the risk of bias introduced when conducting the test.

To create the illusion for participants that they are interacting with a real voice-based system for 3D modeling, we decided to use Blender\footnote{\url{https://www.blender.org}}, explaining to participants that they can interact with it through voice commands. Blender has been selected since it is a free and open-source software that, among other features like sculpting or rendering, allows one to design and visualize 3D objects. One of the main features that made Blender the perfect choice for our \ac{WoZ} study is the availability of \acsp{API} for the Python language\footnote{\url{https://docs.blender.org/api/current/}} that can be used inside a shell-like environment: this allows the Wizard to immediately create and modify the objects programmatically when the participants provide voice commands, thus preventing the participants from noticing anything odd and increasing the speed at which the Wizard is capable of satisfying the participants' requests. Taking advantage of this feature, we pre-defined a set of functions in a Python module to simplify the use of Blender's APIs for the purpose of this study (the module is available in the supplementary materials, sub-folder \textit{python module}).

\begin{figure}[t]
	\centering
	\begin{subfigure}[t]{0.47\linewidth}
		\centering
		\includegraphics[trim={12cm 0 0 8cm}, clip, width=\linewidth]{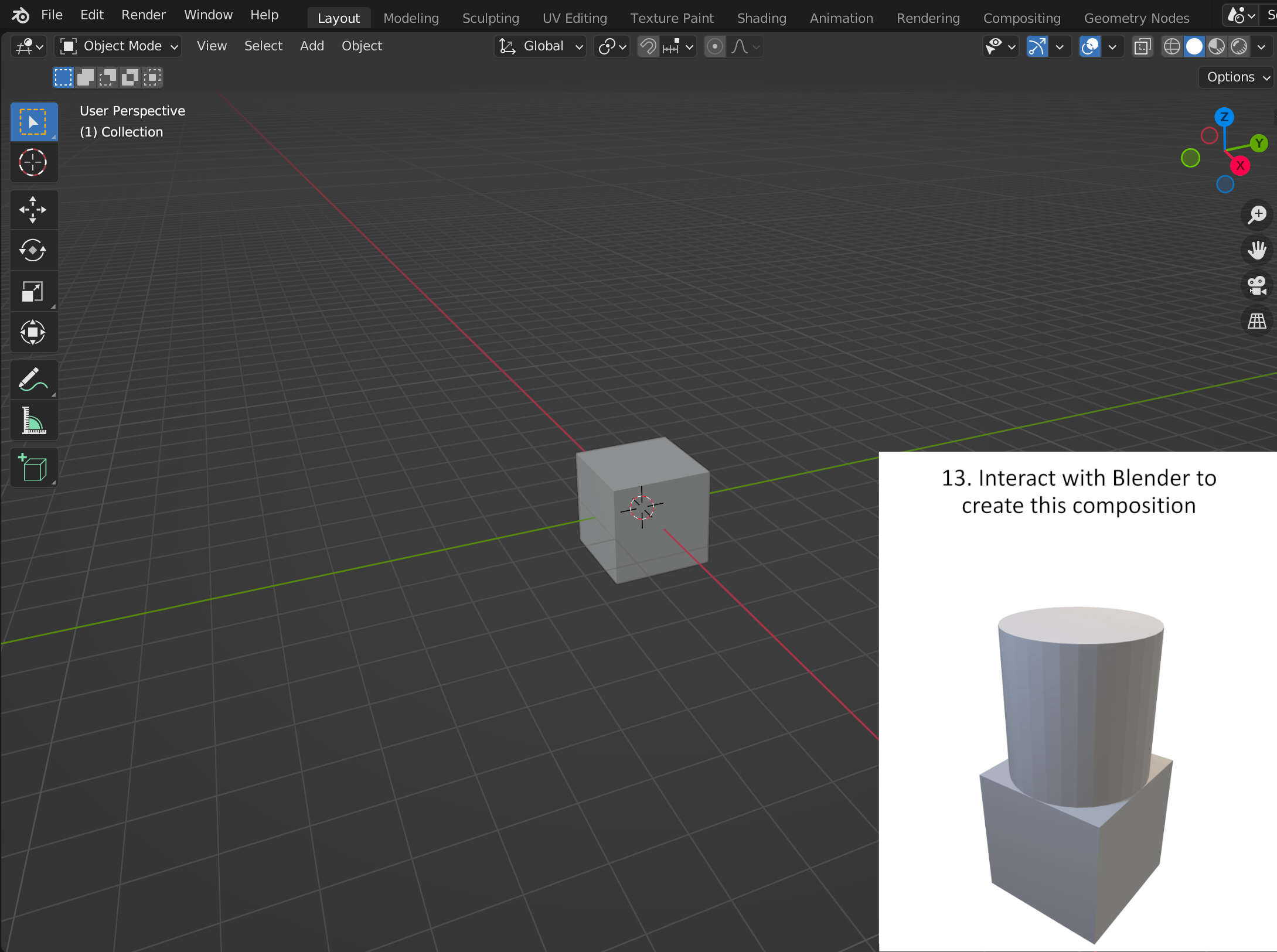}
		\caption{Default screen}
		\label{fig:woz-window-default}
	\end{subfigure}
	\begin{subfigure}[t]{0.47\linewidth}
		\centering
		\includegraphics[trim={12cm 0 0 8cm}, clip, width=\linewidth]{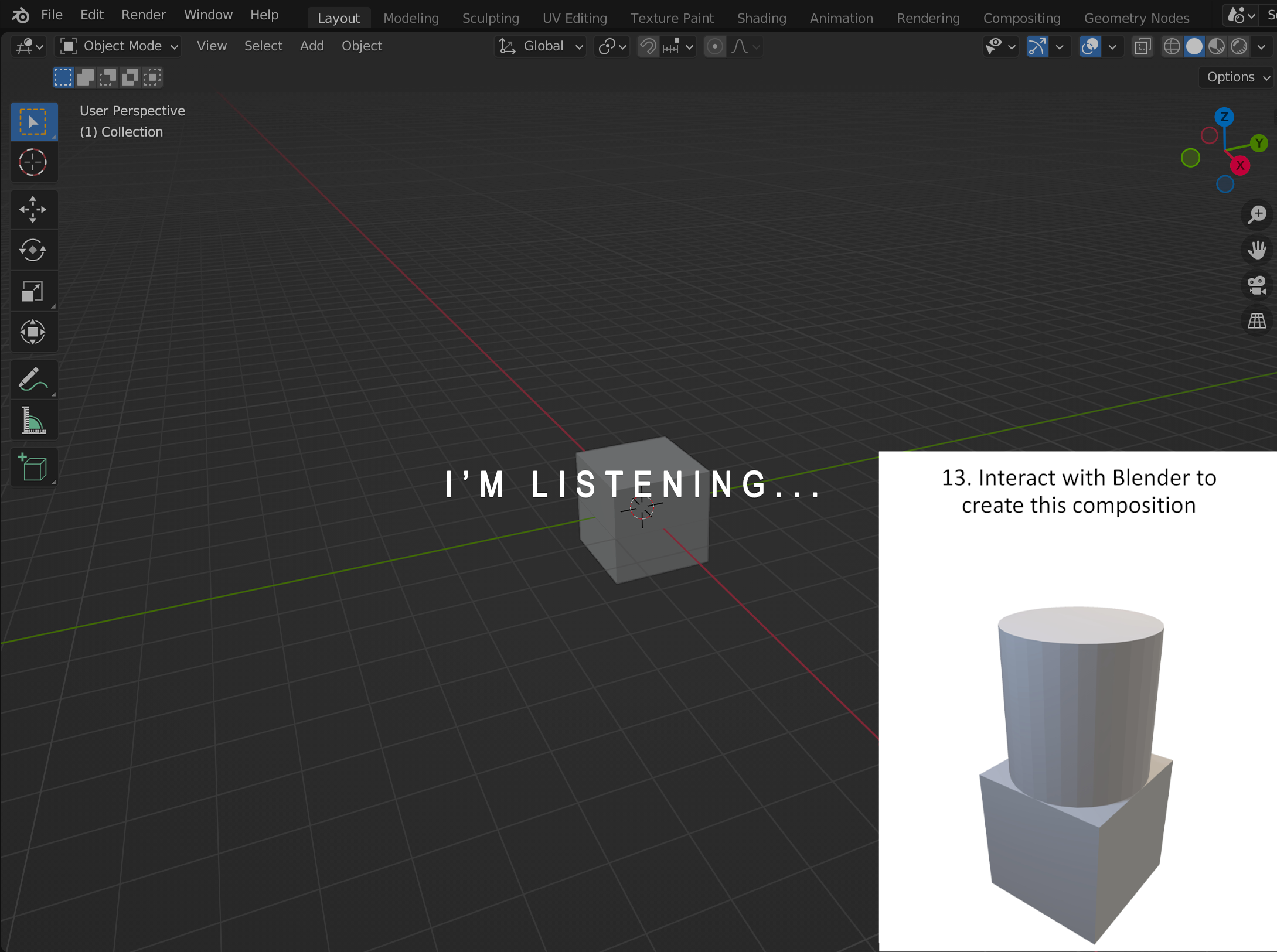}
		\caption{Listening screen}
		\label{fig:woz-window-listening}
	\end{subfigure}\\[\baselineskip] 
	\begin{subfigure}[t]{0.47\linewidth}
		\centering
		\includegraphics[trim={12cm 0 0 8cm}, clip, width=\linewidth]{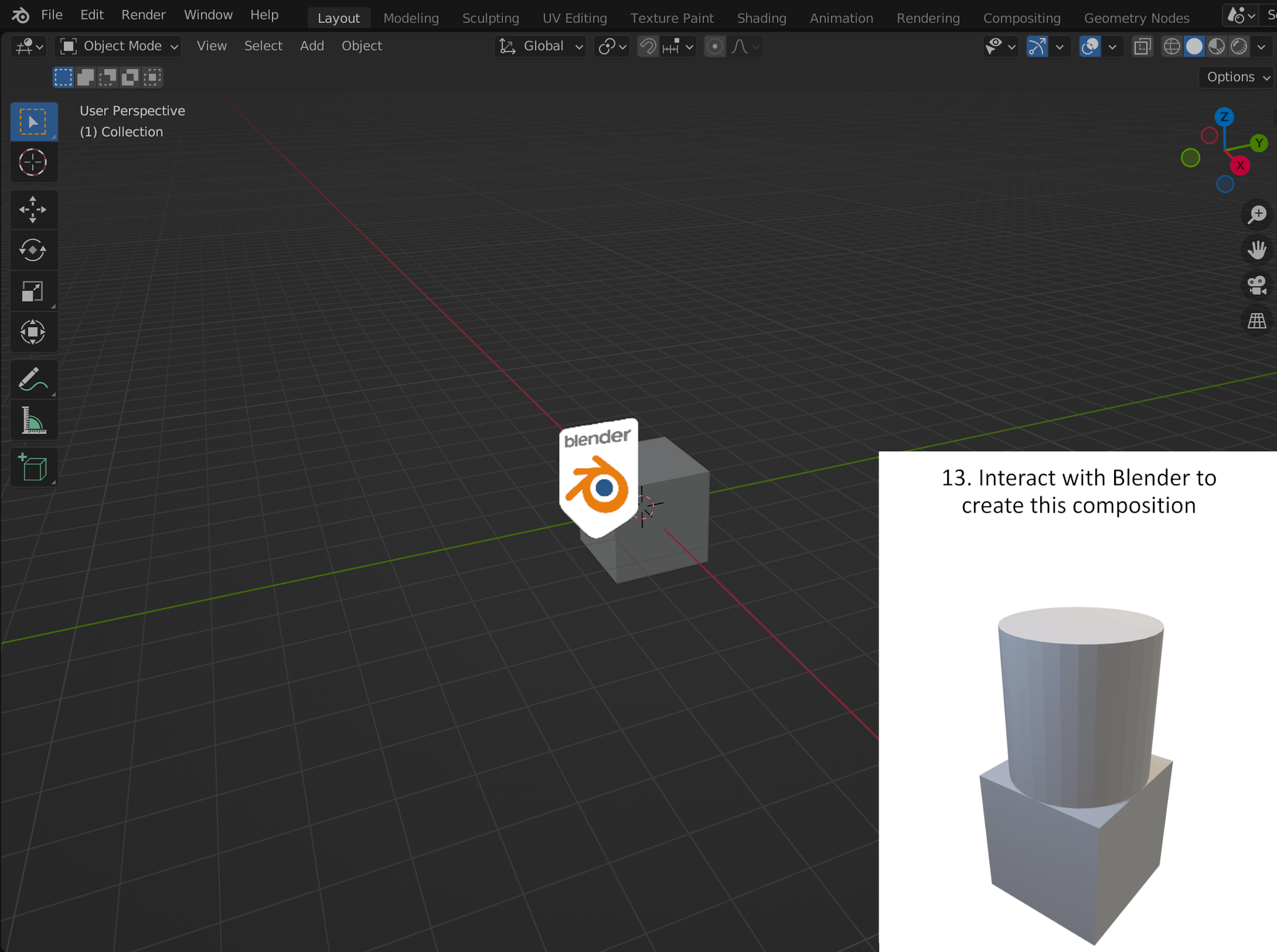}
		\caption{Loading screen}
		\label{fig:woz-window-loading}
	\end{subfigure}
	\begin{subfigure}[t]{0.47\linewidth}
		\centering
		\includegraphics[trim={12cm 0 0 8cm}, clip, width=\linewidth]{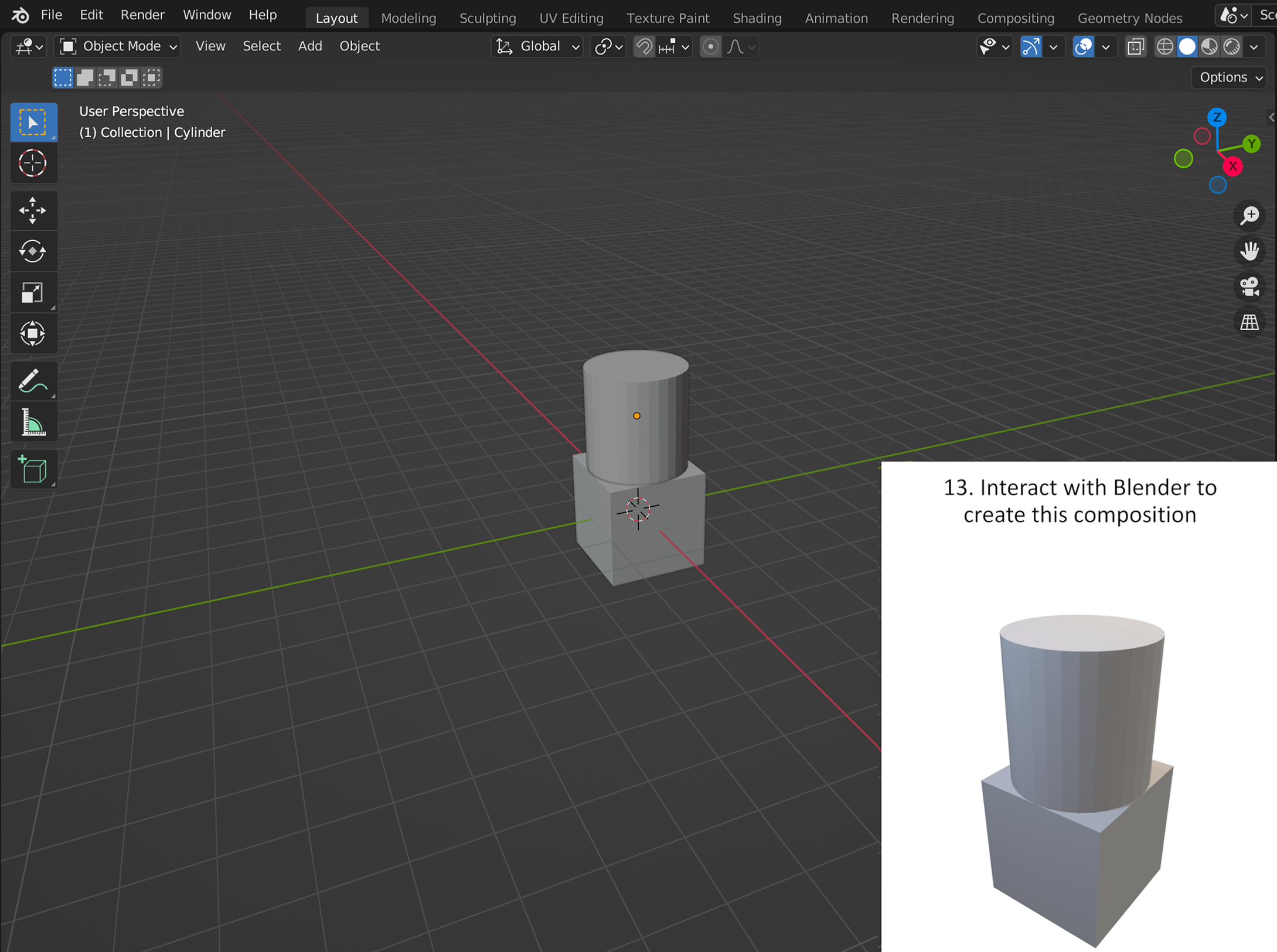}
		\caption{Completed task}
		\label{fig:woz-window-completed}
	\end{subfigure}
	\caption[The window shown to the participants for the \acs{WoZ} study]{The graphical task is  overlaid on the bottom-right side of the Blender’s window from the beginning of the task (\protect\subref{fig:woz-window-default}); when the participants activate the voice assistant by saying "Hey Blender", the “I’m listening” label indicates that they can provide the command to solve the task (\protect\subref{fig:woz-window-listening}); a rotating icon indicates that the voice assistant is elaborating the user commands (\protect\subref{fig:woz-window-loading}); the results is shown after the command elaboration (\protect\subref{fig:woz-window-completed}).}
	\label{fig:woz-window}
\end{figure}

To show the participants the task they had to complete, we overlaid the graphical tasks on the bottom-right side of the Blender's window. To this aim, we used Open Broadcaster Software (or, more commonly, OBS)\footnote{\url{https://obsproject.com}}, a free and open-source software for video recording and live streaming. Using OBS, it was also possible to define animations and transitions to show when users are moving to the next task and to signal to the participants that the ``voice assistant'' (i.e., the Wizard) is listening to the user's command or it is actually performing it. In particular, for each task, both the Blender window and the graphical task are visible (see \cref{fig:woz-window-default}). When the participants activate the Blender voice assistant by saying ``Hey Blender'', the ``I'm listening'' label indicates that participants can provide the command to solve the task (see \cref{fig:woz-window-listening}). Then, when the voice command has been issued, a rotating icon indicates that the voice assistant is analyzing it, creating the illusion that there is a real voice assistant (see \cref{fig:woz-window-loading}). During the loading, the Wizard writes the Python statements related to the user commands and the result is finally shown in Blender (see \cref{fig:woz-window-completed}).

\subsection{Procedure}\label{subsec:procedure}

For each participant, when the Zoom session started, both the conductor and the Wizard were connected on Zoom but the latter never appeared or interacted with the participant. While the conductor introduced the participant to the study, the Wizard shared his screen, in particular the window created by using OBS. The sessions were recorded using Zoom's built-in recorder. Before starting the recordings, participants were asked to sign (either in digital or in verbal form) a privacy policy. It is worth mentioning that our universities require approval by an ethics committee only in the case of medical and clinical studies. For other studies like ours, they require that test participants give consent in a written or digital form; thus, we informed participants about all the details of the study and asked them to agree before starting the study. All of them agreed.

As soon as the participant agreed to attend the study, the conductor invited the participant to complete a set of tasks. The webcam of the conductor was turned off during task execution to avoid disturbing the participant. To reduce the variability between sessions and between the Italian and German participants, the same introductory script was defined (available in the attached appendix, sub-folder "introductory script"). In summary, the conductor explains that the goal of the study was to validate a new voice assistant called Blender, which we created to assist novices in 3D modeling. Then, the conductor asks to complete a set of tasks and that, for each of them, a graphical representation appears on the right-bottom side of their screen. The conductor also specifies that the participant had to first activate the voice assistant by saying ``Hey Blender'' and then, once the ``I'm listening'' label appears, the participant can provide a sequence of voice commands that, in their opinion, is the best to solve the task (for example ``create a cube''). No examples of voice commands have been provided to avoid introducing bias. At the end of each task, the participants had to communicate with the conductor to move on to the next task. 

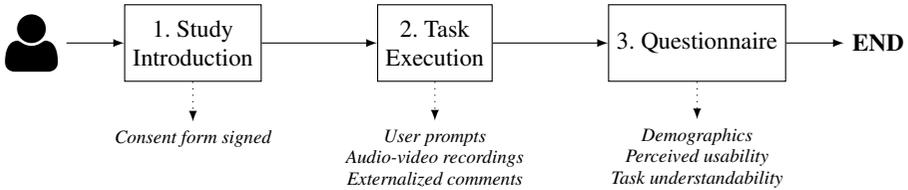
\begin{figure}[h]
\centering
\resizebox{\textwidth}{!}{%
\begin{tikzpicture}[
start chain, node distance=1.5cm,
details/.style={font={\scriptsize\itshape}},
rect/.style={rectangle, minimum height=1cm, text centered, draw=black},
every text node part/.style={align=center}
]
    \node (start) [on chain] {\Huge\faUser};
    \node (intro) [rect, on chain, xshift=-0.75cm] {1. Study \\ Introduction};
    \node (exec) [rect, on chain] {2. Task\\Execution};
    \node (quest) [rect, on chain] {3. Questionnaire};
    \node (end) [on chain, xshift=-0.75cm] {\textbf{\small END}};

    \draw[-latex] (start) -- (intro);
    \draw[-latex] (intro) -- (exec);
    \draw[-latex] (exec) -- (quest);
    \draw[-latex] (quest) -- (end);

    \node (introdet) [details, below of=intro, yshift=0.25cm] {Consent form signed};
    \node (execdet) [details, below of=exec] {User prompts \\ Audio-video recordings \\ Externalized comments};
    \node (questdet) [details, below of=quest] {Demographics\\ Perceived usability\\ Task understandability};

    \draw[-latex, dotted] (intro) -- (introdet);
    \draw[-latex, dotted] (exec) -- (execdet);
    \draw[-latex, dotted] (quest) -- (questdet);
\end{tikzpicture}
}
\caption{Phases of the study and data collected at each phase}
\label{fig:procedure}
\end{figure}

At the end of the session, each participant filled in a  questionnaire that includes questions on demographics, as well as some usability-related questions to evaluate the effectiveness of the Blender voice assistant. Furthermore, since (to the extent of our knowledge) there were no previous examples of graphical tasks for a \acl{WoZ} study, we have also chosen to add some questions to evaluate how easy it was for the user to understand the tasks (available in attached appendix, sub-folder \emph{questionnaire}). The entire procedure lasted around 30 minutes for each participant. A graphical synthesis of the entire procedure and the data collected is shown in \cref{fig:procedure}.

\subsection{Data Analysis}\label{subsec:data}

The first analysis regarded the questionnaire answers that evaluate the choice of providing the tasks in graphical format. Specifically, we included a question that asked ``\emph{How easy it was to understand the graphical tasks?}'' and it ranges from 1 (not simple at all) to 10 (very simple). Both the median and average scores are 8.2/10, with a standard deviation of 1.0. These results seem to validate the idea of presenting the tasks graphically, but it also highlights that for some tasks (the ones with an ambiguous representation) the conductor of the study must be able to guide the participants to the right interpretation (without the use of words that may introduce a primer effect \cite{Etikan2016}). In our study, this issue impacted only the 11th task for four participants and it was solved by turning the webcam on and mimicking the action depicted in the task, in case the user was showing difficulties in understanding a task or if he/she explicitly requested help.

After ensuring the quality of the graphical tasks, we analyzed the qualitative data collected during the study, which helped us answer the research question, i.e., video transcriptions, questionnaire responses and participants' comments. All the video recordings (a total of about 11 hours) were first transcribed and expanded by including the annotations that identify pauses, the start and the end of the processing by the \ac{WoZ}, and eventual errors or over-correction by the \ac{WoZ}. This dataset was completed by reporting the participants comments and the answers to the three open questions we included in the questionnaire:
\begin{enumerate*}[label=\roman*)]
\item What did you like the most about the system used and the interaction with it?
\item What did you like less about the system and the interaction with it? and
\item Would you use a system like Blender to model in 3D? Please motivate your answer.
\end{enumerate*}

This data was analyzed in a systematic qualitative interpretation using Inductive Thematic Analysis \cite{Braun2006}. The initial coding was conducted independently by four researchers, who are co-authors of this article and are experienced in qualitative data analysis: two of them analyzed the Italian results while the other two the German results. The two couples of researchers began with open coding independently. Once all the data was coded, the set of initial codes was further refined by merging the different codes. This first filtering phase allowed us to obtain a set of code groups that capture meaning at a higher level. The identified code groups were then used by each group to extract the main themes. At the end, both the codes and the themes of the two groups were compared to identify similarities and differences. With the exception of some minor differences related to their naming, both the codes and the themes identified by the two couples of researchers were identical in meaning. The final themes that will be presented here derive from a joint naming session carried out by all four researchers. Only a few small differences were identified, and they will be discussed as part of the design implications. The final codes and themes with the relationships among them are available in the attached appendix, sub-folder \emph{Codes and Themes}.

%% file: content/results.tex
\section{Results}\label{sec:results}

The thematic analysis resulted in the description of five themes reported in the following sub-sections. For each theme, significant participant quotes are reported. For the sake of conciseness, we will refer to participants as ``P'' followed by the participant number, and to the \ac{WoZ} system as simply ``system''.

\subsection{Basic Operations}\label{theme:creation-and-manipulation}

This theme frames the strategies of interactions that novices have when they approach the 3D modeling activities of creation and manipulation.

\subsubsection{Creation.}\label{theme:creation}
Novices tend to provide simple commands in the form ``\texttt{<verb> a <shape>}'', where the used verbs are typically ``create'', ``draw'', ``build'', and examples of shape names are ``cube'', ``box'', or ``cylinder''. This behavior has been observed in tasks that required the creation of simple or composite objects. Strictly related to this is the object duplication. Novices usually keep the requests simple by asking them to duplicate a precise object, as P4 did in task 12 when he said ``duplicate the cube''. When the novices, instead, have to face the creation of multiple identical objects, without using the duplication requests (for example, because there was no previous copy in the scene), they simply use a basic creation request by also providing the number of copies: this is clearly exemplified by P5 in task 14 in ``create four cylinders''.

\subsubsection{Manipulation}\label{theme:manipulation}
The manipulation operations used by novices during the study are \emph{translation}, \emph{rotation}, and \emph{scaling}. It is worth mentioning that the manipulation operations require some kind of reference frame to be performed; to this aim, novices often use relative references (for more details see theme \textsc{\nameref{theme:mental-model}} where the references used by the novices are discussed).

In more complex cases, novices provided commands containing both a creation request and an implicit manipulation request, where the manipulation is often expressed as a set of constraints on the final object. As an example, in task 14, P8 asked the system to ``create four cylinders on the corners of the lower rectangle'': in this example, the multiple creation request is clearly visible, and it is put alongside a relative positioning request.

Finally, one of the most interesting identified open codes is the one that relates to moving objects with respect to \emph{implicit construction shapes}. As an example, P4 during the last task asked ``place the four cylinders at the four corners of a square.'' In this example, the participant did not have a square in the scene but implicitly requested the system to create a square, place the cylinders at its corners, and delete the square once the operation was completed. This kind of operation was pretty common throughout the last task: around 45\% of the participants provided a command that used a construction shape like the one previosly cited.

\subsection{Selection of Objects}\label{theme:selection-of-objects}


This theme covers the strategies adopted to identify and select objects, specifically, \emph{absolute} selection, \emph{relative} selection, or \emph{implicit} selection. In the case of absolute selection, most participants explicitly refer to the entire scene, or to a single object in a scene by using its name (the one shown in the ``inspector'' view in Blender, as P11 asked during task 14 by saying ``should I call it Box 0001 if I want to move it?'') or by its shape (as P1 did during task 6 by saying ``move the cube 20 cm downwards''). A specialization of the latter case is the reference to a shape using a 2D approximation. One example is echoed by P8 during task 14: ``Hey blender, move the upper rectangle on the side of the lower one''. Here, the user referred to two 3D boxes by their 2D approximation (rectangles).

The relative selection resulted in four commonly used strategies to select objects, namely:
\begin{itemize}
    \item their relative time of creation (e.g., P3 in task 14: ``Blender, place the second box under the first'');
    \item their relative position (e.g., P8 in task 14: ``Hey Blender, create four cylinders in the corners of the lower rectangle'');
    \item their dimensions (e.g., P11 in task 14: ``Hey Blender, move the tallest box attaching it to the side of the other box'');
    \item by inverting the current selection, eventually applying additional filters (e.g., P3 in task 14: ``Blender, place the other two cylinders like you placed the previous ones'').
\end{itemize}

Finally, users also often performed implicit selections of the objects in the scene, for example, by referring to a single object in the scene or by referring to the last edited object, either explicitly or implicitly (e.g., P1 in task 8 implicitly referred to the last edited object by saying ``increase the volume by three times''). 

It is worth remarking that novices do not differentiate nor have preferences between the various methods, and actually, often mix them to be sure that the selection is clear and precise (e.g.: in a previously shown example by P8 in task 14, ``Hey blender, move the upper rectangle on the side of the lower one'', the user performs the selection by using both an absolute reference to the 2D approximation of the shape of an object, and a relative reference to the positioning of another object).

\subsection{Errors}\label{theme:errors}

Due to the lack of geometry knowledge and/or 3D modeling expertise, often novices commit \emph{errors of which the users are aware of}, and \emph{errors of which the users are not aware of}. In the first case, they try to prevent or correct the errors. For this reason, we named it ``error correction''. In the second case, when a user is either not aware of an error or if they do not care about trying to fix it, then the error simply represents a mistake made during the task execution. For this reason, we named it ``execution errors''. We analyze the details of each thread in the following paragraphs.

\subsubsection{Error correction.}\label{theme:errors-correction}
Different behaviors for correcting the errors have been observed, specifically \emph{during} and \emph{after} the command. Regarding the error correction made during the command, some novices try to prevent their own errors when they recognize one while stating the command, by providing a correction in the same command. For example, P9 during the chair construction task says ``Hey blender, create a rectangle over the quadrilateral of length -- I mean, height 30 centimeters, depth 5 and side 20--22...''. This command contains multiple corrections, starting from the correction of the name of the dimension that the user wants to set to 30 centimeters, and then correcting the actual size of the side of the rectangle to 22 centimeters

Regarding the corrections made after the commands, most of the participants expected some utility commands that are typically available in GUI-based software, like the ``undo'' and ``redo'' functions. As an example, P3 during task 14 provided both the command ``Blender, undo the last operation'', and ``place the other two cylinders as you've placed the previous ones.'' This highlights how, although novices may not be familiar with the task of 3D modeling or voice-based interaction, they were able to transfer the knowledge of other software they may have used in the past, expecting that their previous experience would be applicable to the new, unknown system.

\subsubsection{Execution errors.}\label{theme:execution-errors}
Some of the mistakes committed by the novices are strictly related to \emph{lapsus}, \emph{lack of knowledge}, or \emph{system shortcomings}. In the case of lapsus, some participants referred to shapes and objects using the wrong name (e.g., P10 was trying to refer to a box by calling it ``cylinder'' during task 14). In case of lack of knowledge, errors range from wrong names used for dimensions and primitives, to being unaware of the direction of the axis, perhaps by referring to previous knowledge obtained in school. For example, the Y axis in a 2D plane is usually the vertical one, thus some novices expect the Y axis to be the vertical one also in 3D. Finally, we identified system shortcomings, i.e\@. errors made by the wizard during the execution of the commands: all of these errors can be traced back to the incomprehension of the command, often due to its intrinsic vagueness (see the theme of ``\textsc{\nameref{theme:mental-model}}'').

\subsection{The Gulf of Execution}\label{theme:mental-model}

This theme represents the way novices translate their goals into commands. Throughout the sessions, before providing specific commands, we immediately noticed that novices often think aloud to understand what they have to do and how they can translate it to commands like P16 said during task 14 by saying ``so, the picture has a different point of view. I should move it a little bit. Ok. Hey Blender, make the cylinder bigger.'' Then, by analyzing their commands, we identified three main aspects of the commands where the gulf of execution becomes critical, specifically:
\begin{enumerate*}[label=\roman*)]
    \item relativity
    \item vagueness
     \item abstraction.
\end{enumerate*}

\subsubsection{Relativity.}\label{theme:relativity}
Here we summarize how novices think about positions, scale, rotation, and selection relative to other parts of the scene. Two main overall frames of reference are used by the novices: the axes and other objects. 

To select an axis, novices adopt three approaches, namely:
\begin{enumerate*}[label=\roman*)]
    \item \emph{axis relative direction:} a common way of selecting axes is through their relative direction (depending on the user's point of view), as echoed by P9 during task 11, by saying ``move the geometric shape 20 cm to the right'';
    \item \emph{the axis color:} as an example, during the execution of the last task (the one of creating a chair), P2 referred to the Y axis by its color stating ``turn of 180 degrees the box on the green axis'';
    \item \emph{axis name:} some novices also refer to axes by their actual name, as P19 did during the 12th task by asking the system to ``move the right cube 10 centimeters along the X axis.''.
\end{enumerate*}

When referring to objects' dimensions, novices adopted two main approaches for selection. A first approach consists of using the dimensions' name, as P3 has done in the task of chair creation by saying \emph{``move along the y axis of a length equal to the base of the second box the last cylinder''}. A second approach used a relative comparison to other dimensions; for example, P3 during task 14 selected an object by stating \emph{``move the third cylinder under the highest box [...]''}.

\subsubsection{Vagueness.}\label{theme:vagueness}
It encloses a lack of information in the commands provided to reach the goals. In general, the lack of information is caused by:
\begin{itemize}
    \item \emph{chaining of multiple commands} to describe at a high level a composite shape, as shown by P22 during the chair creation task, by asking ``create four cylinders with the same distance to each other.''; 
    \item \emph{missing data} that the system needs to execute the requests; as an example, novices forget to provide some or all dimensions of a shape (e.g., P1 in task 1 stated ``create a cube'' without providing any dimension), they forget to specify a parameter for a transformation (e.g., P7 in task 10 asked to ``rotate of 30 degrees the figure'' without specifying a direction).
\end{itemize}

\subsubsection{Abstraction.}\label{theme:abstraction}
We noticed two behaviors related to the abstraction of the commands. The first one relates a general abstraction over the process to reach the desired goal, as exemplified by P2 that tried to solve task 14 by saying ``create a chair using two boxes and four cylinders''.  The second one refers to how novices translate the desired 3D shapes into words. For example, shapes are created by providing a general description (e.g., P10 in task 4 by saying ``create a 3D rectangle 30 cm high, 20 cm deep, and long 10 cm'', referred to a box as a ``3D rectangle'', thus simply describing the shape) or by approximating the desired shape with a similar 2D shape (e.g., P8 during task 4 used ``rectangle'' instead of ``box'' by saying ``create a rectangle of height 30, width 20, depth 10''). Furthermore, especially German participants, novices also refer to the 3D shapes by using similar real-world objects (e.g., P17 during task 3 stated ``create a dice with an edge length of 30 centimeters'', using ``dice'' instead of ``cube'').

\subsection{Users' Requests}\label{theme:users-suggestions}

We collected requests and suggestions provided by the participants, which provide useful insights on novices' mental model.

Among the most common requests, participants often asked to rotate the camera and change their point of view. As an example, P11 during the last task of creating a chair, asked ``can I see it from below?'' and ``can I see it from above'' to perform some minor adjustments and corrections to the positions of the 3D objects. This behavior underlines the need to provide a way to allow novices to rotate their point of view. This functional requirement is strictly related to the theme of \textsc{\nameref{theme:selection-of-objects}} as it may benefit from different interaction modalities that could be explored (e.g., using \acl{AR}).

Another common request is related to the actual dimensions: when novices explicitly set size in the command (for example, in the third task), they want to check that the system created an object of the right size. This is exemplified by P10 which explicitly asked if ``can I ask it to check the dimensions?'' in the third task. This suggestion does not translate to an additional requirement for the \ac{AI} model that recognizes users' commands, but it rather provides some insights on the requirements of the whole 3D modeling tool.

Other minor suggestions regarded the customization of the axis: some participants expected the Y axis to be the ``vertical'' one as it usually happens in 2D drawings, rather than the Z axis as it happens in 3D modeling tools like Blender. Providing such a customization option would surely reduce the error rate in a final system, as the novices could adapt it to their own knowledge.

%% file: content/discussion.tex
\section{Discussion and Implications}
\label{sec:discussion}
Based on the findings of the WoZ study, in the following we present design implications for the development of future voice-based 3D modeling tools for novice designers and relate them to the wider research literature around voice assistants and general user experience principles.

\subsubsection{Understand user corrections and adapt to them.} This requirement stems from the errors the users are aware of (see theme \textsc{\nameref{theme:errors}}). It poses requirements that impact two different facets of future voice-based digital modeling tools: the \ac{NLU} layer and the conversation flow. 
Regarding the \ac{NLU} layer, systems must be able to intercept user corrections and aborted commands. Based on our findings, we note that \emph{recognizing uncertainty, hesitation, doubt, and error awareness early on is particularly crucial in the digital modeling context}, as users displayed them frequently due to their unfamiliarity with 3D modeling \cite{Bonner1994}.

Regarding the conversation flow, after intercepting the error correction, it is important to design a dialog that helps users understand the error and recover from it \cite{jain2018evaluating}. Moore and Arar \cite{Moore2019} provide valuable pointers through their \emph{Natural Conversation Framework} which proposes a set of conversational patterns. Some of these patterns relate to \emph{user corrections} and can be applied to voice-based digital modeling. An example inspired by this framework that relates to errors that users correct while they issue a 3D modeling command might be:

\begin{quote}
\textit{User:} Hey blender, increase of 10 centimeters -no- of 20 centimeters the sides of the geometric figure \\
\textit{Agent:} I'm sorry, I didn't understand. Do you mean an increase of 10 or 20 centimeters? \\
\textit{User:} 20 centimeters. \\
\textit{Agent:} Ok, I'm increasing of 20 centimeters the sides of the geometric figure.  \\
\end{quote}

\subsubsection{Deal with vague and incomplete commands}. We have identified numerous \textsc{\nameref{theme:errors}} by the lack of knowledge and the system's shortcomings that users were unaware of. These errors are related to incomprehension due to the vagueness and abstraction of some commands. Self-repair strategies should be introduced to improve interaction \cite{10.1145/3449101}. To this aim, we identified two possible solutions. The first one consists of \emph{sensible defaults}: in case of a vague command, the voice assistant fixes it by \emph{selecting a relevant parameter from a list of alternatives.} For example, if the user says ``create a cylinder on top of the cube'', the cylinder diameter is not specified. In this case, the system can assume that the diameter is equal to the side of the cube. This solution can also benefit from the dialog context: as suggested by Jain et al., \emph{resolving and maintaining the dialog context} can help select the most appropriate sensible default from a list of alternatives \cite{jain2018evaluating}. For example, if other cylinders have been previously created with a given diameter on top of cubes the same can be applied to the new ones in case of vague commands. This allows the system to be proactive, anticipating the users' requests as suggested by V\"olkel et al.~\cite{Volkel2021}.

The second solution consists of \emph{interactively guiding the user} by providing the missing information. With reference to the previous command of the box and cylinder, instead of using defaults, the voice assistant can explicitly ask the user for the desired radius. The strategy adopted by the voice assistant is informed by the degree of system autonomy or desired user control. A hybrid solution can also benefit from both approaches: the selected sensible default can be used by the voice assistant to ask the user if the default is right, for example, with reference to the previous case the voice assistant can reply: ``OK, I'm creating a cylinder with a diameter equal to the side of the cube. Is it OK?''

\subsubsection{Translate interaction conventions to voice-based digital modeling}. Users commonly apply their experience with software applications to other applications or even different domains. As an example, some participants expected to execute ``undo'' or  ``redo'' commands, which are common across applications and domains. This is in line with the traditional Nielsen heuristics of ``user control and freedom'' and  ``consistency and standard'' \cite{Nielsen1994}. The latter states that ``users should not have to wonder whether different words, situations, or actions mean the same thing'', thus the system should ``follow platform and industry conventions'' (from Nielsen \cite{Nielsen1994web}). For this reason, a voice-based 3D modeling system should provide such common operations, like the aforementioned ``undo'' and ``redo'' commands. Further exploration may be required to clearly define and match the set of expected commands to voice-based digital modeling.

\subsubsection{Adopt simple operations even for the creation of composite 3D models}. Based on the theme \textsc{\nameref{theme:creation-and-manipulation}}, we note that most users follow similar and simple approaches even in complex tasks. For example, by analyzing task 13 (which consisted of creating a figure having a cylinder on top of the cube), multiple approaches might be adopted, but novices used only basic operations (creation and translation) to create both a simple cube and a cylinder and then moving the latter on top of the former. This highlights that, although many technical operations may be implemented in voice assistants for digital modeling, it is important to provide novices with simple operations to create and compose 3D objects, rather than prescribing more complex operations like ``extrusion'' and ``insetting'', which are most adequate for skilled users \cite{Nanjundaswamy2013}. 

\subsubsection{Match digital modeling workflows with novices' expectations and experiences from building physical objects}.
Related to the \nameref{theme:creation-and-manipulation}, but by focusing on the last task (that consisted of the creation of a chair), we noticed that the majority of the users started by creating the base cylinders (almost all users started with a phrase like \emph{``create four cylinders}''). This surely provides an interesting insight on how people approach the creation of composite 3D objects. By creating the base cylinders first, users are basically following an approach that starts from the bottom and proceeds upwards. This is not different from the approach that users should follow if they were composing physical shapes: by starting from the bottom, they are able to stack the various shapes without the risk of their composition to ``fall down''. This indication can be useful if wizard procedures are introduced to guide the creation of composite 3D objects; for example, the voice assistants can start the interaction by asking which is the shape, with its features, that must be placed at the bottom, then going on guiding the user to create other shapes on top of the previous ones.

\subsubsection{Provide alternatives for the selection of 3D objects}. By reflecting on the theme of \textsc{\nameref{theme:selection-of-objects}}, we argue that it is among the most critical ones: most of the 3D modeling revolves around the selection of objects to be composed. We found that several and different techniques have been adopted by the novices. For example, a common solution is represented by commands to select an object by referring to the entire scene, in other words in an absolute way. We also documented commands that use relative references, for example, their relative time of creation, their relative position, their dimensions, and by inverting the current selection. The last approach is represented by the implicit selection of the objects in the scene.  These strategies represent different solutions the users can adopt to select a 3D object, and thus the voice assistant should accommodate all of them. To simplify the interaction, future voice assistants can be complemented with additional interaction modalities like gestures or eye tracking, where users could simply point \cite{friedrich2020combining,khan20193d,KHAN2019102847} or gaze \cite{Mayer2020} at the object or surface they want to select. 

\subsubsection{Understand commands that are relative to the user's point of view}. As described in the themes \textsc{\nameref{theme:mental-model}} and \textsc{\nameref{theme:selection-of-objects}}, users often execute commands that are related to their point of view, in particular, to change the camera perspective, to select an axis, and to select a 3D object. In other words, we found that a common way for novices to issue commands is through the ``screen'' coordinate system \cite{Shum1998}, as provided by some professional 3D modeling systems\footnote{\url{https://shorturl.at/fGLRZ}}, by using common words such as ``left'' and ``right'', as P9 did during task 11 with the command ``move the geometric shape 20 cm to the right''. Furthermore, novices often provided commands relative to both their point of view and other objects (as P10 did during task 13: ``insert a cylinder on top of the cube''). This implies that future voice assistants must be equipped with some way of understanding the 3D context into which the command is provided, and they must take into account the user's point of view during the intent-matching process.

\subsubsection{Grant multiple ways to refer to the axes}. Users referred to the axes of the 3D scene by adopting different approaches: by indicating the axis color, by referring to the user's relative direction, by using the axis name (see  themes \textsc{\nameref{theme:mental-model}}) or some users also preferred to switch the Y and Z axes as the ``vertical'' axis (see theme \textsc{\nameref{theme:users-suggestions}}). This ambiguity is also found in professional systems, as some of them use the Z axis as vertical while others use the Y axis instead \cite{van2011blender}. This behavior should be considered in the design of voice assistants for 3D modeling, since this is a core activity that, if not adequately supported, might lead to ineffective user interaction. 

\subsubsection{Design for complex commands.}. Multiple chained commands have often been prompted to execute various actions. In our study, it was possible to accommodate the multiple users commands thanks to the \ac{WoZ} but voice assistants are typically restricted to simple standalone commands. Similar to what Fast et al\@. already proposed for complex tasks \cite{FastCMBB17}, also voice-based systems for 3D modeling should address this requirement, which strongly impacts the design of its \ac{NLU} layer that must be able to understand and execute multiple chained commands.

\subsubsection{Favor explicit trigger words}. Previous work by Vtyurina et al\@. argued that forcing the use of explicit trigger words would constrain user interactions, suggesting the use of implicit conversation cues for driving the dialog \cite{Vtyurina2018}. On the contrary, during our experiments novices used implicit conversational cues while thinking about their workflow and as a natural reaction after a successful command execution (see \textsc{\nameref{theme:mental-model}}): this highlights the need for future voice-based systems to provide clear explicit activation cues and trigger words, to avoid any unintentional activation that would disrupt users' workflow.

\subsubsection{Embrace diversity in naming approaches}. As novices usually have little to no knowledge of the 3D modeling domain, they often have to resort to different naming approaches when dealing with shapes for which they do not recall the ``right'' name. As already highlighted in \textsc{\nameref{theme:mental-model}}, novices can refer to shapes by providing high-level descriptions (e.g., ``3D rectangle'' instead of ``box''), 2D approximations (``rectangle'' instead of ``box''), or by associating them to a real-world object (e.g., ``dice'' instead of ``cube''). For this reason, future systems must be able to understand both analogies and descriptions of shapes. A concrete solution might be the adoption of a lexical ontology like WordNet \cite{WordNet} to infer the shape name related to the real object.


%% file: content/limitations.tex
\section{Limitations of the Study}
\label{sec:limitations}
Our study is an initial step toward understanding how novices approach voice-based 3D modeling. We have identified some limitations of our work. First, the novices' languages deserve a wider exploration: our study highlights very small differences between Germans and Italians because of their culture; however, a similar study where participants use their native languages might be useful to understand how language might impact the resulting mental model. Similarly, this study does not focus on how aspects like ethnicity, socio-economic status, and age might impact the novice's mental model. Another limitation regards the tasks: the ones used in the study are representative of the most common operations to design 3D models but digital fabrication often implies the design of objects that are more complex than a chair. In addition, the set of proposed tasks does not cover all possible operations (e.g., selecting textures and making holes). Future work may also study differences between the mental model of lay users (target of this study) and novices in 3D modeling that are domain experts (e.g., they have expertise in sculpting or 3D world composition, but do not know how to model). Similarly, the proposed voice-based interaction approach may be compared with alternative solutions based on mouse and keyboard or multi-modal approaches, to explore the pros and cons of each solution. Finally, Blender has been selected as the 3D modeling tool because of the advantages reported in \cref{subsec:apparatus}; however, its \acs{UI} is designed for a \acs{WIMP} interaction thus it presents commands, buttons, functions, etc., that might bias or confuse novices. Despite carefully hiding all the useless parts of the Blender \acs{UI}, the adoption of a system purposely designed to better fit the voice interaction might be adopted to elicit the mental model.

%% file: content/conclusion.tex
\section{Conclusion}
\label{sec:conclusion}

Voice interaction is emerging as a promising paradigm that can simplify 3D modeling for digital fabrication. However, novices' mental model is never considered when designing voice-based 3D modeling systems. In addition, voice interaction is usually built on top of \acs{WIMP} systems instead of designing the voice paradigm and the whole system from scratch. This study addresses these limitations by investigating the novices' mental model in 3D modeling and contributes to the state-of-the-art by identifying a set of design implications that support the definition of voice-based interaction paradigms for the design and customization of personalized 3D models. This contribution aims to lower the barrier to 3D modeling thus supporting the wider democratization of digital fabrication.

As future work, we are now addressing the limitations reported in the previous section. We are also working on the development of a prototype of a voice assistant integrated into Blender: it is currently being developed in DialogFlow \cite{Sabharwal2020} and it has been designed considering the design implications proposed in this study. The aim is to study novices' behavior when interacting with real systems, also exploring if and how the design indications suggested in this study also accommodate the design of more complex objects in more realistic situations, for example, by proposing scenarios instead of tasks.

%% file: content/acknowledgments.tex
This work has been funded by the European Union’s Horizon 2020 research and innovation program under grant agreement No. 952026 (\url{https://www.humane-ai.eu/}).
The research of Andrea Esposito is funded by a Ph.D. fellowship within the framework of the Italian ``D.M. n. 352, April 9, 2022'' - under the National Recovery and Resilience Plan, Mission 4, Component 2, Investment 3.3 - Ph.D. Project ``Human-Centered Artificial Intelligence (HCAI) techniques for supporting end users interacting with AI systems'', co-supported by ``Eusoft S.r.l.'' (CUP H91I22000410007).

%% file: interact2023-digital-modeling-for-everyone.bbl
\begin{thebibliography}{10}
\providecommand{\url}[1]{\texttt{#1}}
\providecommand{\urlprefix}{URL }
\providecommand{\doi}[1]{https://doi.org/#1}

\bibitem{Billinghurst1997}
Billinghurst, M., Baldis, S., Matheson, L., Philips, M.: 3d palette: A virtual
  reality content creation tool. In: Proceedings of the ACM Symposium on
  Virtual Reality Software and Technology. pp. 155--156. VRST '97, Association
  for Computing Machinery, Lausanne, Switzerland (1997).
  \doi{10.1145/261135.261163}

\bibitem{Bonner1994}
Bonner, S.E.: A model of the effects of audit task complexity. Accounting,
  Organizations and Society  \textbf{19}(3),  213--234 (1994).
  \doi{10.1016/0361-3682(94)90033-7}

\bibitem{Braun2019}
Braun, M., Mainz, A., Chadowitz, R., Pfleging, B., Alt, F.: At your service:
  Designing voice assistant personalities to improve automotive user
  interfaces. In: Proceedings of the 2019 CHI Conference on Human Factors in
  Computing Systems. pp. 1--11. CHI '19, Association for Computing Machinery,
  New York, NY, USA (2019). \doi{10.1145/3290605.3300270}

\bibitem{Braun2006}
Braun, V., Clarke, V.: Using thematic analysis in psychology. Qualitative
  Research in Psychology  \textbf{3}(2),  77--101 (2006).
  \doi{10.1191/1478088706qp063oa}

\bibitem{Cuadra2021}
Cuadra, A., Goedicke, D., Zamfirescu-Pereira, J.: Democratizing design and
  fabrication using speech: Exploring co-design with a voice assistant. In: CUI
  2021 - 3rd Conference on Conversational User Interfaces. CUI '21, Association
  for Computing Machinery, Bilbao (online), Spain (2021).
  \doi{10.1145/3469595.3469624}

\bibitem{10.1145/3449101}
Cuadra, A., Li, S., Lee, H., Cho, J., Ju, W.: My bad! repairing intelligent
  voice assistant errors improves interaction. Proc. ACM Hum.-Comput. Interact.
   \textbf{5}(CSCW1) (apr 2021). \doi{10.1145/3449101}

\bibitem{davis1952automatic}
Davis, K.H., Biddulph, R., Balashek, S.: Automatic recognition of spoken
  digits. The Journal of the Acoustical Society of America  \textbf{24}(6),
  637--642 (1952). \doi{10.1121/1.1906946}

\bibitem{Etikan2016}
Etikan, I., Musa, S.A., Alkassim, R.S.: Comparison of convenience sampling and
  purposive sampling. American Journal of Theoretical and Applied Statistics
  \textbf{5}(1), ~1--4 (01 2016). \doi{10.11648/j.ajtas.20160501.11}

\bibitem{FastCMBB17}
Fast, E., Chen, B., Mendelsohn, J., Bassen, J., Bernstein, M.S.: Iris: {A}
  conversational agent for complex tasks. CoRR  \textbf{abs/1707.05015} (2017),
  \url{http://arxiv.org/abs/1707.05015}

\bibitem{Feeman2018}
Feeman, S.M., Wright, L.B., Salmon, J.L.: Exploration and evaluation of cad
  modeling in virtual reality. Computer-Aided Design and Applications
  \textbf{15}(6),  892--904 (2018). \doi{10.1080/16864360.2018.1462570}

\bibitem{Fialho2015}
Fialho, P., Coheur, L.: Chatwoz: Chatting through a wizard of oz. In:
  Proceedings of the 17th International ACM SIGACCESS Conference on Computers
  \& Accessibility. pp. 423--424. ASSETS '15, Association for Computing
  Machinery, Lisbon, Portugal (2015). \doi{10.1145/2700648.2811334}

\bibitem{friedrich2020combining}
Friedrich, M., Langer, S., Frey, F.: Combining gesture and voice control for
  mid-air manipulation of cad models in vr environments. In: Proceedings of the
  16th International Joint Conference on Computer Vision, Imaging and Computer
  Graphics Theory and Applications - Volume 1: HUCAPP,. pp. 119--127. INSTICC,
  SciTePress, Online Streaming (2021). \doi{10.5220/0010170501190127}

\bibitem{gershenfeld2012make}
Gershenfeld, N.: How to make almost anything: The digital fabrication
  revolution. Foreign Affairs  \textbf{91}, ~43 (2012)

\bibitem{grigorvoice}
Grigor, S., Nandra, C., Gorgan, D.: Voice-controlled 3d modelling with an
  intelligent personal assistant. International Joural of User-System
  Interaction  \textbf{13},  73--88 (01 2020). \doi{10.37789/ijusi.2020.13.2.2}

\bibitem{GRIMES2021113515}
Grimes, G.M., Schuetzler, R.M., Giboney, J.S.: Mental models and expectation
  violations in conversational ai interactions. Decision Support Systems
  \textbf{144},  113515 (2021). \doi{10.1016/j.dss.2021.113515}

\bibitem{van2011blender}
van Gumster, J.: Blender For Dummies. For Dummies, USA, 2nd edn. (2011)

\bibitem{Huang2017}
Huang, Y.C., Chen, K.L.: Brain-computer interfaces (bci) based 3d
  computer-aided design (cad): To improve the efficiency of 3d modeling for new
  users. In: Schmorrow, D.D., Fidopiastis, C.M. (eds.) Augmented Cognition.
  Enhancing Cognition and Behavior in Complex Human Environments. pp. 333--344.
  Springer International Publishing, Cham (2017)

\bibitem{jain2018evaluating}
Jain, M., Kumar, P., Kota, R., Patel, S.N.: Evaluating and informing the design
  of chatbots. In: Proceedings of the 2018 Designing Interactive Systems
  Conference. pp. 895--906. DIS '18, Association for Computing Machinery, New
  York, NY, USA (2018). \doi{10.1145/3196709.3196735}

\bibitem{James2018}
James, J., Watson, C.I., MacDonald, B.: Artificial empathy in social robots: An
  analysis of emotions in speech. In: Proceedings of the 27th IEEE
  International Symposium on Robot and Human Interactive Communication
  (RO-MAN). pp. 632--637. RO-MAN 2018, IEEE Press, Nanjing, China (2018).
  \doi{10.1109/ROMAN.2018.8525652}

\bibitem{JOWERS2013923}
Jowers, I., Prats, M., McKay, A., Garner, S.: Evaluating an eye tracking
  interface for a two-dimensional sketch editor. Computer-Aided Design
  \textbf{45}(5),  923--936 (2013). \doi{10.1016/j.cad.2013.01.006}

\bibitem{KHAN2019102847}
Khan, S., Tun{\c c}er, B.: Gesture and speech elicitation for 3d cad modeling
  in conceptual design. Automation in Construction  \textbf{106},  102847
  (2019). \doi{10.1016/j.autcon.2019.102847}

\bibitem{khan20193d}
Khan, S., Tuncer, B., Subramanian, R., Blessing, L.: 3d cad modeling using
  gestures and speech: Investigating cad legacy and non-legacy procedures. In:
  Lee, J.H. (ed.) Proceedings of the 18th International Conference on Computer
  Aided Architectural Design Futures. pp. 347--366. CAAD Futures 2019,
  CUMINCAD, Daejeon, Republic of Korea (2019),
  \url{http://papers.cumincad.org/cgi-bin/works/paper/cf2019_042}

\bibitem{Kou2008}
Kou, X.Y., Tan, S.T.: Design by talking with computers. Computer-Aided Design
  and Applications  \textbf{5}(1-4),  266--277 (2008).
  \doi{10.3722/cadaps.2008.266-277}

\bibitem{KOU2010545}
Kou, X., Xue, S., Tan, S.: Knowledge-guided inference for voice-enabled cad.
  Computer-Aided Design  \textbf{42}(6),  545--557 (2010).
  \doi{10.1016/j.cad.2010.02.002}

\bibitem{Langevin2021}
Langevin, R., Lordon, R.J., Avrahami, T., Cowan, B.R., Hirsch, T., Hsieh, G.:
  Heuristic evaluation of conversational agents. In: Proceedings of the 2021
  CHI Conference on Human Factors in Computing Systems. CHI '21, Association
  for Computing Machinery, New York, NY, USA (2021).
  \doi{10.1145/3411764.3445312}

\bibitem{Lee2008}
Lee, M., Billinghurst, M.: A wizard of oz study for an ar multimodal interface.
  In: Proceedings of the 10th International Conference on Multimodal
  Interfaces. pp. 249--256. ICMI '08, Association for Computing Machinery,
  Chania, Crete, Greece (2008). \doi{10.1145/1452392.1452444}

\bibitem{Mayer2020}
Mayer, S., Laput, G., Harrison, C.: Enhancing mobile voice assistants with
  worldgaze. In: Proceedings of the 2020 CHI Conference on Human Factors in
  Computing Systems. pp. 1--10. CHI '20, Association for Computing Machinery,
  Honolulu, HI, USA (2020). \doi{10.1145/3313831.3376479}

\bibitem{medhi2017you}
Medhi~Thies, I., Menon, N., Magapu, S., Subramony, M., O'Neill, J.: How do you
  want your chatbot? an exploratory wizard-of-oz study with young, urban
  indians. In: Bernhaupt, R., Dalvi, G., Joshi, A., K.~Balkrishan, D., O'Neill,
  J., Winckler, M. (eds.) Human-Computer Interaction - INTERACT 2017. pp.
  441--459. Springer International Publishing, Cham (2017)

\bibitem{WordNet}
Miller, G.A.: Wordnet: A lexical database for english. Commun. ACM
  \textbf{38}(11),  39--41 (nov 1995). \doi{10.1145/219717.219748}

\bibitem{Moore2019}
Moore, R.J., Arar, R.: Conversational UX Design: A Practitioner's Guide to the
  Natural Conversation Framework. Association for Computing Machinery, New
  York, NY, USA (2019)

\bibitem{Murad2019}
Murad, C., Munteanu, C., Cowan, B.R., Clark, L.: Revolution or evolution?
  speech interaction and hci design guidelines. IEEE Pervasive Computing
  \textbf{18}(2),  33--45 (2019). \doi{10.1109/MPRV.2019.2906991}

\bibitem{murad2021finding}
Murad, C., Munteanu, C., R.~Cowan, B., Clark, L.: Finding a new voice:
  Transitioning designers from gui to vui design. In: CUI 2021 - 3rd Conference
  on Conversational User Interfaces. CUI '21, Association for Computing
  Machinery, New York, NY, USA (2021). \doi{10.1145/3469595.3469617}

\bibitem{Nanjundaswamy2013}
Nanjundaswamy, V.G., Kulkarni, A., Chen, Z., Jaiswal, P., S., S.S., Verma, A.,
  Rai, R.: Intuitive {3D} computer-aided design ({CAD}) system with multimodal
  interfaces. In: Proceedings of the ASME 2013 International Design Engineering
  Technical Conferences and Computers and Information in Engineering
  Conference. International Design Engineering Technical Conferences and
  Computers and Information in Engineering Conference, vol. Volume 2A: 33rd
  Computers and Information in Engineering Conference. ASME, Portland, Oregon,
  USA (08 2013). \doi{10.1115/DETC2013-12277}, v02AT02A037

\bibitem{Nielsen1994web}
Nielsen, J.: 10 usability heuristics for user interface design (1994),
  \url{https://www.nngroup.com/articles/ten-usability-heuristics/}

\bibitem{Nielsen1994}
Nielsen, J.: Enhancing the explanatory power of usability heuristics. In:
  Proceedings of the SIGCHI Conference on Human Factors in Computing Systems.
  pp. 152--158. CHI '94, Association for Computing Machinery, Boston,
  Massachusetts, USA (1994). \doi{10.1145/191666.191729}

\bibitem{niu2022multimodal}
Niu, H., Van~Leeuwen, C., Hao, J., Wang, G., Lachmann, T.: Multimodal natural
  human--computer interfaces for computer-aided design: A review paper. Applied
  Sciences  \textbf{12}(13), ~6510 (2022)

\bibitem{nowacki2020improving}
Nowacki, C., Gordeeva, A., Liz{\'e}, A.H.: Improving the usability of voice
  user interfaces: A new set of ergonomic criteria. In: Marcus, A., Rosenzweig,
  E. (eds.) Design, User Experience, and Usability. Design for Contemporary
  Interactive Environments. pp. 117--133. Springer International Publishing,
  Cham (2020)

\bibitem{plumed2021voice}
Plumed, R., Gonz{\'a}lez-Lluch, C., P{\'e}rez-L{\'o}pez, D., Contero, M.,
  Camba, J.D.: {A voice-based annotation system for collaborative
  computer-aided design}. Journal of Computational Design and Engineering
  \textbf{8}(2),  536--546 (4 2021). \doi{10.1093/jcde/qwaa092}

\bibitem{Sabharwal2020}
Sabharwal, N., Agrawal, A.: Introduction to Google Dialogflow, pp. 13--54.
  Apress, Berkeley, CA (2020). \doi{10.1007/978-1-4842-5741-8\_2}

\bibitem{Samad1985}
Samad, T., Director, S.W.: Towards a natural language interface for cad. In:
  Proceedings of the 22nd ACM/IEEE Design Automation Conference. pp.~2--8. DAC
  '85, IEEE Press, Las Vegas, Nevada, USA (1985)

\bibitem{seaborn2021voice}
Seaborn, K., Miyake, N.P., Pennefather, P., Otake-Matsuura, M.: Voice in
  human--agent interaction: a survey. ACM Computing Surveys (CSUR)
  \textbf{54}(4),  1--43 (2021)

\bibitem{Sharp2019}
Sharp, H., Rogers, Y., Preece, J.: Interaction Design: Beyond Human Computer
  Interaction. John Wiley \& Sons, Inc., Hoboken, NJ, USA (2007)

\bibitem{Shum1998}
Shum, H.Y., Han, M., Szeliski, R.: Interactive construction of 3d models from
  panoramic mosaics. In: Proceedings. 1998 IEEE Computer Society Conference on
  Computer Vision and Pattern Recognition (Cat. No.98CB36231). pp. 427--433.
  IEEE, Santa Barbara, California (1998). \doi{10.1109/CVPR.1998.698641}

\bibitem{SREESHANKAR201451}
{Sree Shankar}, S., Rai, R.: Human factors study on the usage of bci headset
  for 3d cad modeling. Computer-Aided Design  \textbf{54},  51--55 (2014).
  \doi{10.1016/j.cad.2014.01.006}, aPPLICATION OF BRAIN--COMPUTER INTERFACES IN
  CAD/E SYSTEMS

\bibitem{STARK2010179}
Stark, R., Israel, J., W{\"o}hler, T.: Towards hybrid modelling
  environments---merging desktop-cad and virtual reality-technologies. CIRP
  Annals  \textbf{59}(1),  179--182 (2010). \doi{10.1016/j.cirp.2010.03.102}

\bibitem{Thakur2015}
Thakur, A., Rai, R.: {User Study of Hand Gestures for Gesture Based 3D CAD
  Modeling}. In: Proceedings of the ASME 2015 International Design Engineering
  Technical Conferences and Computers and Information in Engineering
  Conference. International Design Engineering Technical Conferences and
  Computers and Information in Engineering Conference, vol. Volume 1B: 35th
  Computers and Information in Engineering Conference. ASME, Boston,
  Massachusetts, USA (08 2015). \doi{10.1115/DETC2015-46086}, v01BT02A017

\bibitem{Voice2CAD}
{Voice2CAD}: {Voice2CAD} (2022), \url{https://voice2cad.com/}

\bibitem{Volkel2021}
V\"{o}lkel, S.T., Buschek, D., Eiband, M., Cowan, B.R., Hussmann, H.: Eliciting
  and analysing users' envisioned dialogues with perfect voice assistants. In:
  Proceedings of the 2021 CHI Conference on Human Factors in Computing Systems.
  CHI '21, Association for Computing Machinery, Yokohama, Japan (2021).
  \doi{10.1145/3411764.3445536}

\bibitem{Vtyurina2018}
Vtyurina, A., Fourney, A.: Exploring the role of conversational cues in guided
  task support with virtual assistants. In: Proceedings of the 2018 CHI
  Conference on Human Factors in Computing Systems. pp.~1--7. CHI '18,
  Association for Computing Machinery, Montreal QC, Canada (2018).
  \doi{10.1145/3173574.3173782}

\bibitem{VULETIC2021102609}
Vuletic, T., Duffy, A., McTeague, C., Hay, L., Brisco, R., Campbell, G.,
  Grealy, M.: A novel user-based gesture vocabulary for conceptual design.
  International Journal of Human-Computer Studies  \textbf{150},  102609
  (2021). \doi{10.1016/j.ijhcs.2021.102609}

\bibitem{Making_A_Scene}
Vyas, S., Chen, T.J., Mohanty, R., Krishnamurthy, V.: {Making-A-Scene: A
  Preliminary Case Study on Speech-based 3D Shape Exploration through Scene
  Modeling}. Journal of Computing and Information Science in Engineering pp.
  1--11 (08 2022). \doi{10.1115/1.4055239}

\bibitem{xue2009natural}
Xue, S., Kou, X., Tan, S.: Natural voice-enabled cad: modeling via natural
  discourse. Computer-Aided Design and Applications  \textbf{6}(1),  125--136
  (2009)

\bibitem{xue2010command}
Xue, S., Kou, X., Tan, S.: Command search for cad system. Computer-Aided Design
  and Applications  \textbf{7}(6),  899--910 (2010)

\bibitem{Yoon2008}
Yoon, S.M., Graf, H.: Eye tracking based interaction with 3d reconstructed
  objects. In: Proceedings of the 16th ACM International Conference on
  Multimedia. pp. 841--844. MM '08, Association for Computing Machinery,
  Vancouver, British Columbia, Canada (2008). \doi{10.1145/1459359.1459501}

\end{thebibliography}
